\documentclass[preprint,showpacs,amsmath,amssymb,pra,superscriptaddress]{revtex4}

\usepackage{graphicx}
\usepackage{dcolumn}
\usepackage{bm}

\begin{document}


\title{Spectral Properties and Lifetimes of Neutral Spin-$\frac{1}{2}$-Fermions in a Magnetic Guide}
\date{\today}

\author{Igor Lesanovsky}
\email[]{ilesanov@physi.uni-heidelberg.de}
\affiliation{%
Physikalisches Institut, Universit\"at Heidelberg, Philosophenweg 12, 69120 Heidelberg, Germany}%

\author{Peter Schmelcher}
\email[]{Peter.Schmelcher@pci.uni-heidelberg.de}\thanks{Corresponding
author}
\affiliation{%
Physikalisches Institut, Universit\"at Heidelberg, Philosophenweg 12, 69120 Heidelberg, Germany}%
\affiliation{%
Theoretische Chemie, Institut f\"ur Physikalische Chemie,
Universit\"at Heidelberg,
INF 229, 69120 Heidelberg, Germany}%

\date{\today}

\begin{abstract}\label{txt:abstract}
We investigate the resonant motion of neutral
spin-$\frac{1}{2}$-fermions in a magnetic guide. A wealth of
unitary and anti-unitary symmetries is revealed in particular
giving rise to a two-fold degeneracy of the energy levels. To
compute the energies and decay widths of a large number of
resonances the complex scaling method is employed. We discuss the
dependence of the lifetimes on the angular momentum of the
resonance states. In this context the existence of so-called
quasi-bound states is shown. In order to approximately calculate
the resonance energies of such states a radial Schr\"odinger
equation is derived which improves the well-known adiabatic
approximation. The effects of an additionally applied homogeneous
Ioffe field on the resonance energies and decay widths are also
considered. The results are applied to the case of the
$^6\text{Li}$ atom in the $F=\frac{1}{2}$ hyperfine ground state.
\end{abstract}

\maketitle
%
\section{Introduction}\label{sec:introduction}
%

Magnetic traps are now being used for many years in order to study
and manipulate the properties of neutral atoms \cite{Folman02}.
With the availability of efficient cooling techniques the
controlled occupation of low lying quantum levels inside such
traps became possible. This paved the way for the experimental
exploration of several phenomena such as
Bose-Einstein-condensation (see Ref. \cite{Pethick02} and
references therein) or the formation of degenerate Fermi gases
\cite{Schreck01}. Apart from studying such collective properties
there is also a fundamental and vivid interest in the properties
and the dynamics of single atoms. Its understanding is essential
in order to identify systems for quantum information processing
but it is also necessary to gain insights into trapping mechanisms
in order to control the atomic motion. Theoretical studies of the
quantum properties of neutral atoms in magnetic traps have been
performed since the late eighties. Since then several field
configurations have been in the focus of numerous publications,
e.g. the quadrupole field \cite{Bergeman89}, the wire trap
\cite{Hau95,Berg-Soerensen96,Burke96} or the magnetic guide and
the Ioffe trap \cite{Hinds00,Hinds00E,Potvliege01}. However, only
a few field configurations allow for stationary solutions. Bound
states of spin-$\frac{1}{2}$-particles trapped by a wire have been
found analytically by Bl\"umel \emph{et al.} \cite{Bluemell91} and
Vestergaard Hau \emph{et al.} \cite{Hau95}. The latter authors
employed a super-symmetric approach in order to derive the Rydberg
series of bound states around an infinitely thin wire. Finite wire
sizes were accounted for by introducing a quantum defect in the
Rydberg formula. Burke \emph{et al.} \cite{Burke96} were able to
numerically derive similar series of energies for particles with
spin up to 2 by using the finite element method combined with the
application of multichannel quantum defect theory. However, in
most cases atoms are trapped in meta-stable states which have a
finite lifetime. The calculation of the corresponding resonance
energies and lifetimes requires the application of sophisticated
numerical methods. Bergeman \emph{et al.} \cite{Bergeman89} have
calculated dozens of resonances of spin-$\frac{1}{2}$-particles
inside a three-dimensional magnetic quadrupole field by
determining the phase-shift of scattered waves. A similar approach
was pursued by Hinds and Eberlein \cite{Hinds00,Hinds00E} which
have studied the dynamics of neutral particles in a magnetic
guide. Their results were reproduced by Potvliege and Zehnl\'{e}
who have performed a complex scaling calculation. In both
publications the effect of an additionally applied homogeneous
Ioffe field is also discussed but only a few resonances are
studied. Although there are no principle obstacles to compute
resonances there has always been put great effort in finding
reasonable approximations which allow for an quick and easy
calculation of the resonance energies. Here the most prominent
example is the so-called adiabatic approximation
\cite{Bergeman89,Berg-Soerensen96} where the spin is aligned along
the local direction of the magnetic field. Depending on their
electronic state the atoms are here divided into low- and
high-field seekers which give rise to bound and unbound solutions.

In this work we focus on the quantum dynamics of a neutral
spin-$\frac{1}{2}$-particle in a magnetic guide. We also address
the case of an additional homogeneous field being applied parallel
to the guide (Ioffe field). Like Potvliege and
Zehnl\'{e}\cite{Potvliege01} we employ a complex scaling approach.
In contrast to their work we calculate hundreds of resonances for
several values of the Ioffe field strength which enables us to
analyze global features of the resonance spectrum. In order get
further insights we develop an approximate description of
so-called quasi-bound states. We derive a radial Schr\"odinger
equation which is superior to the common adiabatic approach. In
contrast to the latter our method becomes exact for high angular
momenta and its quality does not depend on whether there is an
additional Ioffe field or not.

This work is organized as follows: In Section \ref{txt:system} we
derive the Hamiltonian for a neutral particle possessing a
magnetic moment $\vec{\mu}$ in a magnetic guide. We also account
for an eventually applied homogeneous Ioffe field parallel to the
guide. The resulting two-dimensional Hamiltonian depends on a
single parameter involving both the gradient of the inhomogeneous
field and the Ioffe field strength. It exhibits a plethora of
unitary as well as anti-unitary symmetries. The effects of this
intricate symmetry properties and in particular a resultant
two-fold degeneracy of the resonance energies are discussed in
Sec. \ref{txt:symmetries}. In Sec. \ref{txt:numerics} we briefly
outline the numerical approach we pursue in order to obtain the
resonance energies and decay width which is based on the complex
scaling method. Sec. \ref{txt:results} is devoted to a discussion
of our results. We analyze the resonance spectrum for several
values of the Ioffe field strength. Furthermore we investigate how
the energies and lifetimes of the resonance states depend on their
angular momentum. The properties of so-called quasi bound states
are addressed. We provide a radial Schr\"odinger equation whose
eigenenergies are very good approximations of the true resonance
energies. A comparison to the commonly used adiabatic
approximation is also performed. Finally the results are applied
to the case of $^6\text{Li}$ in a magnetic field generated by a
current carrying wire together with a homogeneous bias field.

%
\section{The Hamiltonian}\label{txt:system}
%
We consider the motion of a neutral spin-$\frac{1}{2}$-particle
which carries the magnetic moment $\vec{\mu}$ inside a magnetic
quadrupole guide. The magnetic field is given by
\begin{eqnarray}
  \vec{B}\left(\vec{r}\right)=\left(%
\begin{array}{c}
  b\,x \\
  -b\,y \\
  B \\
\end{array}%
\right).\label{eq:sideguide_field}
\end{eqnarray}
The parameters $b$ and $B$ determine the gradient and the field
strength of an eventually applied homogeneous field along the
$z$-axis (Ioffe field), respectively. Generally the Hamiltonian
for a neutral particle with the magnetic moment $\vec{\mu}$ and
mass $M$ in a magnetic field reads
\begin{eqnarray}
  H_{3D}=\frac{1}{2M}\left(p_x^2+p_y^2+p_z^2\right)-\vec{\mu}\cdot\vec{B}\left(\vec{r}\right)\label{eq:initial_hamiltonian}
\end{eqnarray}
Since there is in our case (\ref{eq:initial_hamiltonian}) no
explicit dependence on the $z$-coordinate the momentum in
$z$-direction $p_z$ is conserved. Employing plane waves for the
wavefunction in $z$-direction one obtains a two-dimensional
Hamiltonian which describes the dynamics in the $x-y$ plane.
Adopting atomic units \footnote{$\hbar=1$, $m_e=1$, $a_0=1$,
$e=1$: The magnetic gradient unit then becomes
$b=1a.u.=4.44181\cdot 10^{15} \frac{T}{m}$. The magnetic field
strength unit is $B=1a.u.=2.35051\cdot 10^{5} T$} and with
$\vec{\mu}=-\frac{g}{2} \vec{S}$ where $\vec{S}$ is the spin
operator one yields
\begin{eqnarray}
H_{2D}=\frac{1}{2M}\left[p_x^2+p_y^2+\frac{g M}{2}
\left(b\,x\sigma_x-b\,y\sigma_y+B\,\sigma_z\right)\right].
\end{eqnarray}
Here $g$ is the $g$-factor of the particle. The components of
$\vec{S}$ are given through the Pauli matrices $\sigma_i$.
Performing the scale transformation $\bar{x}_i=\left(\frac{b
gM}{2}\right)^\frac{1}{3}x_i$ and $\bar{p}=\left(\frac{b
gM}{2}\right)^{-\frac{1}{3}}p_i$ and omitting the bars one obtains
\begin{eqnarray}
   M\left(\frac{b
gM}{2}\right)^{-\frac{2}{3}}H_{2D}=H=\frac{1}{2}\left(p_x^2+p_y^2+
x\sigma_x-y\sigma_y+\gamma\,\sigma_z\right)
\label{eq:scaled_Hamiltonian}
\end{eqnarray}
with $\gamma=B\left(\frac{g M}{2 b^2}\right)^{\frac{1}{3}}$. For
$\gamma=0$ the gradient $b$ does not explicitly appear in equation
(\ref{eq:scaled_Hamiltonian}). Therefore we find here the energy
level spacing to scale according to $\frac{1}{M}\left(\frac{b
gM}{2}\right)^{\frac{2}{3}}$.

%
\section{Symmetries}\label{txt:symmetries}
%

%
\subsection{The case $\gamma=0$}
%
Analyzing the structure of the Hamiltonian
(\ref{eq:scaled_Hamiltonian}) for $\gamma=0$ we find 15 discrete
symmetry operations. They are listed in table
\ref{tbl:symmetries}.
\begin{table}[htb]\center
\begin{tabular}{|c|c|c|c|}

\hline $\Sigma_x=\sigma_xP_y$& $\Sigma_y=P_x\sigma_y$& $\Sigma_z=P_xP_y\sigma_z$& \\

\hline $I_{xy}S_1$& $P_yI_{xy}S_2$& $P_xP_yI_{xy}S_1^*$&$P_xI_{xy}S_2^*$ \\

\hline\hline $T\sigma_x$& $TP_xP_y\sigma_y$& $TP_x\sigma_z$&$TP_y$\\

\hline $TP_yI_{xy}S_1$&$TI_{xy}S_2$& $TP_xI_{xy}S_1^*$&
$TP_xP_yI_{xy}S_2^*$\\\hline

\end{tabular}\caption{Symmetry operations of the Hamiltonian
(\ref{eq:scaled_Hamiltonian}) for $\gamma=0$. Top part: unitary
symmetries. Bottom part: anti-unitary
symmetries.}\label{tbl:symmetries}
\end{table}
Each symmetry is composed of a number of elementary operations
which are provided in table \ref{tbl:operations}.
\begin{table}[htb]\center
\begin{tabular}{|c||c|}
\hline Operator& Operation\\

\hline\hline $P_{x_i}$& $x_i\rightarrow -x_i$\\

\hline $T$& $A \rightarrow A^*$\\

\hline $\sigma_x$ &$\sigma_y\rightarrow
-\sigma_y$\quad$\sigma_z\rightarrow
-\sigma_z$\\

\hline $\sigma_y$ &$\sigma_x\rightarrow
-\sigma_x$\quad$\sigma_z\rightarrow
-\sigma_z$\\

\hline $\sigma_z$ &$\sigma_x\rightarrow
-\sigma_x$\quad$\sigma_y\rightarrow
-\sigma_y$\\

\hline $I_{xy}$& $x \rightarrow y$\quad $y \rightarrow x$\quad\quad($\phi \rightarrow -\phi+\frac{\pi}{2}$)\\

\hline $S_1=\left(%
\begin{array}{cc}
  0 & 1 \\
  -i & 0 \\
\end{array}%
\right)$& $\sigma_x\rightarrow -\sigma_y$\quad$\sigma_y\rightarrow -\sigma_x$\quad $\sigma_z\rightarrow -\sigma_z$ \\

\hline $S_2=\left(%
\begin{array}{cc}
  -i & 0 \\
  0 & 1 \\
\end{array}%
\right)$& $\sigma_x\rightarrow -\sigma_y$\quad$\sigma_y\rightarrow
\sigma_x$\quad $\sigma_z\rightarrow \sigma_z$
\\\hline
\end{tabular}\caption{Set of discrete operations out of which
 all symmetry operations of the Hamiltonian (\ref{eq:scaled_Hamiltonian}) can be composed.}\label{tbl:operations}
\end{table}
All symmetry operations are either unitary or anti-unitary, the
anti-unitary ones involving the conventional time reversal
operator $T$. In spite of its simplicity the system
(\ref{eq:scaled_Hamiltonian}) possesses a wealth of symmetry
properties. The algebra of the underlying symmetry group possesses
a complicated structure some features of which will be discussed
in the following. The operators $\Sigma_x$, $\Sigma_y$ and
$\Sigma_z$ generate a sub-group obeying the algebra
$\left[\Sigma_i,\Sigma_j\right]=2i\,\epsilon_{ijk}\Sigma_k$
reminiscent of angular momentum operators. In addition one finds
$\Sigma_i^2=1$. Interestingly these quantities act on both real
and spin space.

Apart from the discrete symmetries there is a continuous symmetry
namely $\Lambda_z=L_z-S_z$ which was already pointed out by
Eberlein and Hinds in Ref. \cite{Hinds00E}. The operator
$\Lambda_z$ obeys the eigenvalue equation
\begin{eqnarray}
\Lambda_z\left|m\right>=m\left|m\right>
\end{eqnarray}
with the half-integer quantum numbers $m$. For
spin-$\frac{1}{2}$-particles the states $\left|m\right>$ read in
the spatial representation
\begin{eqnarray}
  \left|m\right>=\left(%
\begin{array}{c}
  \alpha^{\uparrow}\,e^{i(m+\frac{1}{2})\phi} \\
  \alpha^{\downarrow}\, e^{i(m-\frac{1}{2})\phi} \\
\end{array}%
\right).
\end{eqnarray}
The symmetries listed in table \ref{tbl:symmetries} together with
$\Lambda_z$ allow several sets of commuting operators. For our
investigation we choose the set consisting of $H$, $\Sigma_z$ and
$\Lambda_z$. Calculating the eigenvalue of a $\Lambda_z$
eigenstate with respect to the $\Sigma_z$ operator leads to
\begin{eqnarray}
\Sigma_z\left|m\right>=\left(%
\begin{array}{c}
 \alpha^{\uparrow}\, e^{i(m+\frac{1}{2})(\phi+\pi)} \\
  -\alpha^{\downarrow}\,e^{i(m-\frac{1}{2})(\phi+\pi)} \\
\end{array}%
\right)=e^{i(m+\frac{1}{2})\pi}\left|m\right>=\kappa\left|m\right>.
\label{eq:kappa_m_mapping}
\end{eqnarray}
Therefore in case of $m+\frac{1}{2}$ being odd we find $\kappa=-1$
whereas we obtain $\kappa=1$ for even values of $m+\frac{1}{2}$.

Exploiting the discrete symmetries one can show that there is a
two-fold degeneracy of any energy level. This degeneracy is
revealed as follows. The operations $\Sigma_z$ and $P_y\sigma_x$
obey $\left\{\Sigma_z,P_y\sigma_x\right\}=0$. Let
$\left|E,\kappa\right>$ be an energy eigenstate and at the same
time an eigenstate of $\Sigma_z$ with
\begin{eqnarray}
\Sigma_z\left|E,\kappa\right> &=& \kappa\left|E,\kappa\right>
\end{eqnarray}
and $\kappa=\pm 1$. Employing the above anti-commutator one
obtains
\begin{eqnarray}
\Sigma_z P_y\sigma_x \left|E,\kappa\right> = -P_y\sigma_x
\Sigma_z\left|E,\kappa\right>= -\kappa P_y\sigma_x
\left|E,\kappa\right>
\end{eqnarray}
The state $P_y\sigma_x \left|E,\kappa\right>$ can be identified
with $\left|E,-\kappa\right>$. Hence, as long as $\kappa\neq 0$
\footnote{Since $\Sigma_z$ is a unitary operator the case
$\kappa=0$ cannot occur.} there is always an orthogonal pair of
states possessing the same energy namely $\left|E,\kappa\right>$
and $\left|E,-\kappa\right>$. We have to emphasize that there
occur no further degeneracies in the system. In principle one
could think of performing the above calculation repeatedly but now
substituting $P_y\sigma_x$ with any operator listed in table
\ref{tbl:symmetries} which anti-commutes with $\Sigma_z$. It turns
out that all of the resulting states generated by this scheme are
either superpositions of $\left|E,\kappa\right>$ and
$\left|E,-\kappa\right>$ or differ only by a phase factor from one
of these states.

Since $\Sigma_z$ and $\Lambda_z$ commute, i.e.
$\left[\Sigma_z,\Lambda_z\right]=0$, one can construct eigenstates
with respect to both of these operators:
$\left|E,m,\kappa\right>$. Letting $P_y\sigma_x$ act on these
states together with considering the anti-commutator
$\left\{P_y\sigma_x,\Lambda_z\right\}=0$ one finds
\begin{eqnarray}
\Lambda_z\,
P_y\sigma_x\left|E,m,\kappa\right>=-P_y\sigma_x\,\Lambda_z\left|E,m,\kappa\right>=-m\,P_y\sigma_x\left|E,m,\kappa\right>.
\end{eqnarray}
Hence we can identify $P_y\sigma_x\left|E,m,\kappa\right>$ with
$\left|E,-m,\kappa\right>$. Therefore the pairs of degenerate
states do not only have opposite $\kappa$ quantum number but also
opposite $m$-values. A detailed discussion of degeneracies of
spin-$\frac{1}{2}$-particles can be found in \cite{Lesanovsky04}.

%
\subsection{The case $\gamma\neq 0$}
%
For $\gamma\neq 0$ the conserved quantity $\Lambda_z$ persists but
only 7 of the discrete symmetries for $\gamma=0$ remain. They are
listed in table \ref{tbl:symmetries_gamma_neq_0}.
\begin{table}
\begin{tabular}{|c|c|c|c|}

\hline $\Sigma_z$&$P_yI_{xy}S_2$&$P_xI_{xy}S_2^*$&\\

\hline\hline $TP_y$&$TP_x\sigma_z$ &$TP_xP_yI_{xy}S_2^*$ & $TI_{xy}S_2$\\

\hline

\end{tabular}\caption{Discrete symmetry operations of the Hamiltonian
(\ref{eq:scaled_Hamiltonian}) for a finite Ioffe field strength
$B$, i.e. $\gamma\neq 0$. Top part: unitary symmetries. Bottom
part: anti-unitary symmetries.
symmetries.}\label{tbl:symmetries_gamma_neq_0}
\end{table}
These operations form a non-Abelian algebra. In contrast to the
group operations listed in table \ref{tbl:symmetries} there are no
two anti-commuting operators. Hence it is not possible to construct
pairs of degenerate energy eigenstates as discussed above. The
operation $P_y\sigma_x$ which we have used to construct the
degenerate pairs of states for $\gamma=0$ now has the following
property
\begin{eqnarray}
  P_y\sigma_x\,H\left(\gamma\right)\,P_y\sigma_x=H\left(-\gamma\right),
\end{eqnarray}
i.e. the sign of $\gamma$ in the Hamiltonian
(\ref{eq:scaled_Hamiltonian}) is changed. Hence, the application
of $P_y\sigma_x$ reverses the sign of the homogeneous Ioffe field.
We emphasize that even for $\gamma\neq 0$ the operators $H$,
$\Sigma_z$ and $\Lambda_z$ form a commuting set of operators.

%
\section{Numerical Treatment}\label{txt:numerics}
%

The Hamiltonian (\ref{eq:scaled_Hamiltonian}) does not support
bound states \cite{Hinds00E}. In order to calculate the energies
and decay widths of the scattering wavefunctions we employ the
complex scaling method together with the linear variational
principle.

We perform a rotation of the spatial coordinates into the complex
plane: $x_i\rightarrow x_i e^{i \theta}$ which yields the
'rotated' Hamiltonian
\begin{eqnarray}
H\left(\theta\right)=\frac{1}{2}\left(p_x^2 e^{-i 2 \theta
}+p_y^2e^{-i 2\theta }+ xe^{i \theta }\sigma_x-ye^{i \theta
}\sigma_y\right).\label{eq:complex_scaled_Hamiltonian}
\end{eqnarray}
Unlike the bound states only continuum states are affected by the
complex scaling procedure. The scattering states are rotated into
the lower half of the complex plane and in particular resonances
correspond now to square integrable functions. Since for
$\theta\neq 0$ the complex scaled Hamiltonian
(\ref{eq:complex_scaled_Hamiltonian}) is not Hermitian one
encounters complex energies:
\begin{eqnarray}
  \varepsilon=E-i\frac{\Gamma}{2}.
\end{eqnarray}
Here $E$ is the energy and $\Gamma$ the decay width of the
resonance (the lifetime $\tau$ of a quantum states is connected to
its decay width by $\tau=\Gamma^{-1}$). In general the energy $E$
will depend on the value of the rotation angle $\theta$. Continuum
states are rotated by $2 \theta$ into the lower half of the
complex plane whereas energies of resonance states - once revealed
- are independent of the angle $\theta$. A detailed discussion of
the complex scaling procedure can be found in Ref.
\cite{Moiseyev98} and Refs. therein. Since the complex scaling
procedure transforms resonance states into square integrable
wavefunctions one can apply the linear variational principle in
order to compute the wavefunctions and resonance energies. We
utilize an orthonormal basis set of the form
\begin{eqnarray}
\left|m,n,s\right>=\left|m,n\right>\left|s\right>
\end{eqnarray}
where the functions $\left|n,m\right>$ are products of harmonic
oscillator eigenfunctions
\begin{eqnarray}
\left|m,n\right>=\frac{\sqrt[4]{\omega_x
\omega_y}}{\sqrt{2^{m+n}\pi\,
m!\,n!}}\,e^{-\frac{1}{2}\left(\omega_x x^2+\omega_y y^2\right)
}H_m\left(\sqrt{\omega_x}x\right)H_n\left(\sqrt{\omega_y}y\right)
\label{eq:basis_set}
\end{eqnarray}
The parameters $\omega_x$ and $\omega_y$ are non-linear
variational parameters which can be adapted in order to gain an
optimal convergence behavior of the numerically obtained
wavefunctions. To cover the spin space dynamics we utilize the
spinor-orbitals $\left|s\right>=\left|\downarrow\right>$ and
$\left|\uparrow\right>$, respectively.  Performing the linear
variational principle the solutions $\left|E\right>$ of the
stationary Schr\"odinger equation are expanded in a finite set of
the functions (\ref{eq:basis_set}):
\begin{eqnarray}
  \left|E\right>=\sum_{nms}c_{mn}\left|m,n,s\right>
\end{eqnarray}
From our knowledge of the symmetry properties we can choose a more
specific appearance of this expansion. Requiring the states
$\left|E\right>$ to be eigenstates with respect to $\Sigma_z$ we
make use of the properties
\begin{eqnarray}
\Sigma_z\left|m,n,\uparrow\right>&=&\left(-1\right)^{m+n}\left|m,n,\uparrow\right>\\
\Sigma_z\left|m,n,\downarrow\right>&=&\left(-1\right)^{m+n+1}\left|m,n,\downarrow\right>.
\end{eqnarray}
Thus the expansions for the two $\kappa$ subspaces read
\begin{eqnarray}
 \left|E,\kappa=+1\right>&=&\sum_{m+n=\text{even}}a_{mn}\left|m,n,\uparrow\right>+\sum_{m+n=\text{odd}}b_{mn}\left|m,n,\downarrow\right>\\
 \left|E,\kappa=-1\right>&=&\sum_{m+n=\text{odd}}a_{mn}\left|m,n,\uparrow\right>+\sum_{m+n=\text{even}}b_{mn}\left|m,n,\downarrow\right>.
\end{eqnarray}
Since $\left[\Lambda_z,\Sigma_z\right]=0$ one could also demand
the states to be a priori eigenfunctions of $\Lambda_z$ as well.
However, our basis set is not particularly well suited to
construct eigenfunctions to $\Lambda_z$. We therefore abstain from
putting this constraint on the above expansion. Nevertheless, the
resulting diagonalization of the Hamiltonian matrix (see below)
guarantees that our numerically obtained energy and
$\Sigma_z$-eigenfunctions are also eigenfunctions to $\Lambda_z$.
\begin{figure}[htb]\center
\includegraphics[angle=0,width=6.5cm]{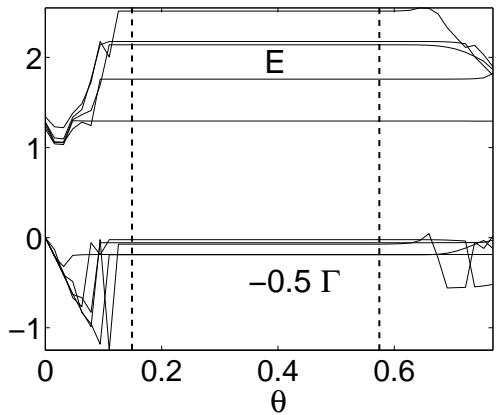}
\caption{Dependence of the resonance energies (upper part) and
decay widths (lower part) on the complex scaling angle $\theta$ as
observed in the numerical calulations. The stationary region is
located in between the dashed lines.} \label{fig:theta_dependence}
\end{figure}
The determination of the optimal expansion coefficients $a_{mn}$
and $b_{mn}$ gives rise to a complex algebraic eigenvalue problem
$\mathbf{H}\left(\theta\right)\vec{v}\left(\theta\right)=\varepsilon\left(\theta\right)\vec{v}\left(\theta\right)$,
where $\mathbf{H}$ is the matrix representation of the Hamiltonian
(\ref{eq:complex_scaled_Hamiltonian}). The vector $\vec{v}$
contains the coefficients $a_{mn}$ and $b_{mn}$. The eigenvalues
of $\mathbf{H}\left(\theta\right)$ are computed for
$0\leq\theta\leq\frac{\pi}{4}$. Due to the finite size of the
employed basis set resonances become not independent of the
complex scaling angle $\theta$ but exhibit a stationary behavior
within a certain $\theta$-interval as can be seen in figure
\ref{fig:theta_dependence}.
%
\section{Results}\label{txt:results}
%
In this section we discuss the results obtained via the
above-described complex scaling approach. We present the resonance
energy spectrum and study the dependence of the decay widths on
the eigenvalue of $\Lambda_z$. Additionally we discuss how the
lifetimes are affected if a homogeneous Ioffe field ist applied.
Furthermore a radial Schr\"odinger equation is derived that
describes quasi-bound states in the magnetic guide. All results
are given for $b=M=\frac{g}{2}= 1$. In order to obtain the
resonance energies and decay widths for different parameters one
has to multiply the latter by the factor $\frac{1}{M}\left(\frac{b
g}{2}\right)^{\frac{2}{3}}$. The section concludes with a
discussion of how the results apply to the experimentally
important case of a trapped $^6 \text{Li}$ atom in the
$F=\frac{1}{2}$ hyperfine ground state.
%
\subsection{Energies and decay widths of resonance states}\label{txt:resonances}
\begin{figure}[htb]\center
\includegraphics[angle=0,width=7cm]{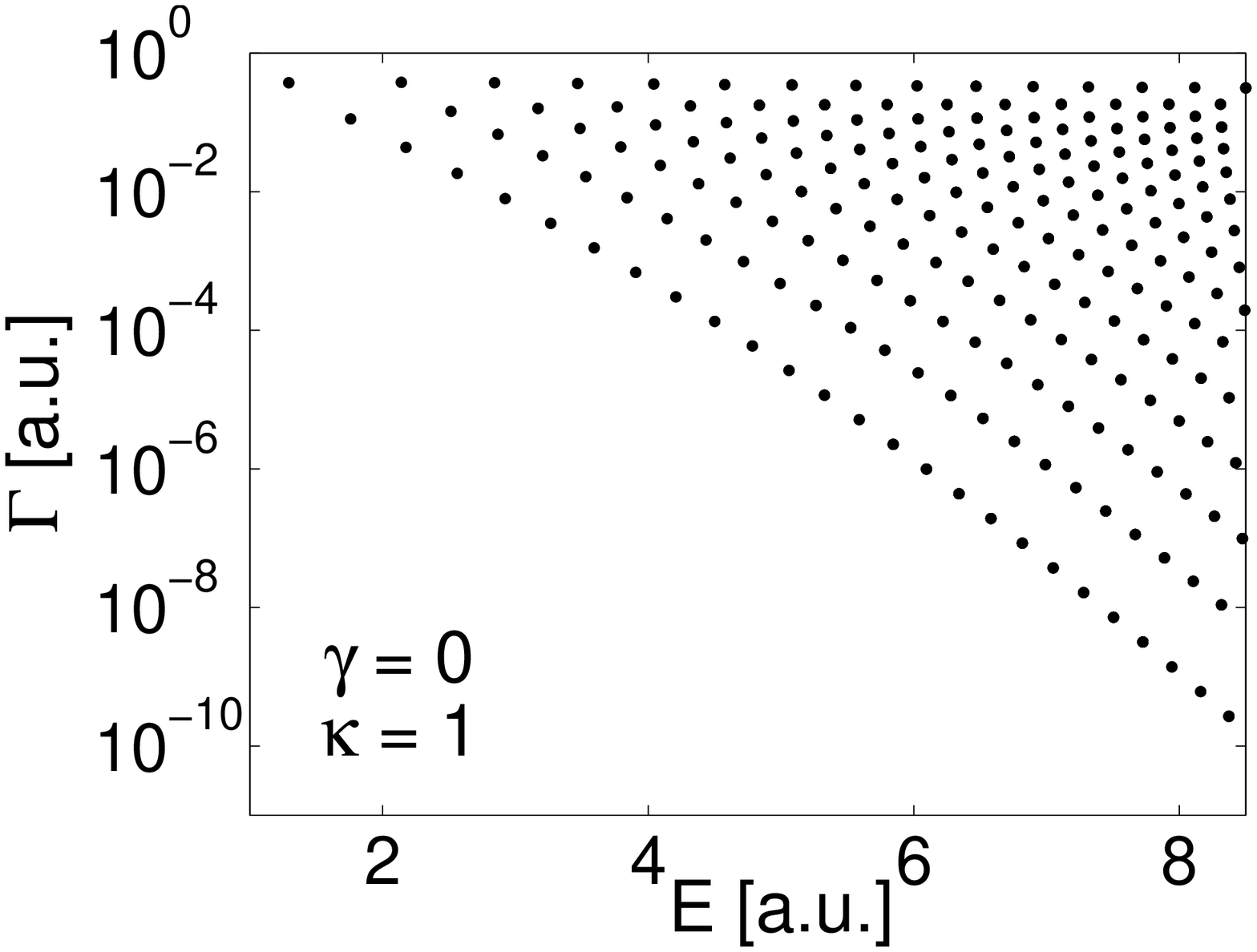}\includegraphics[angle=0,width=7cm]{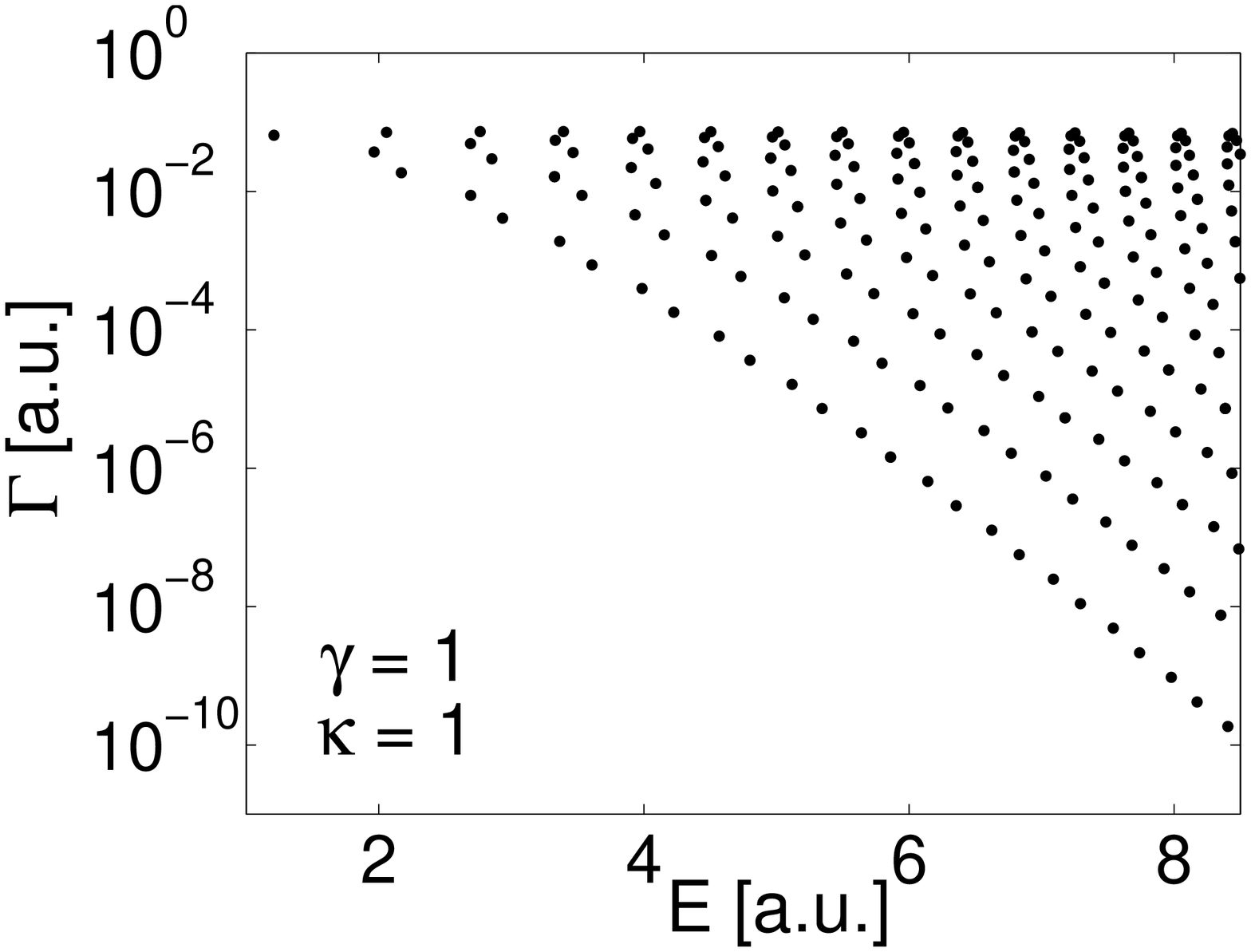}\\
\includegraphics[angle=0,width=7cm]{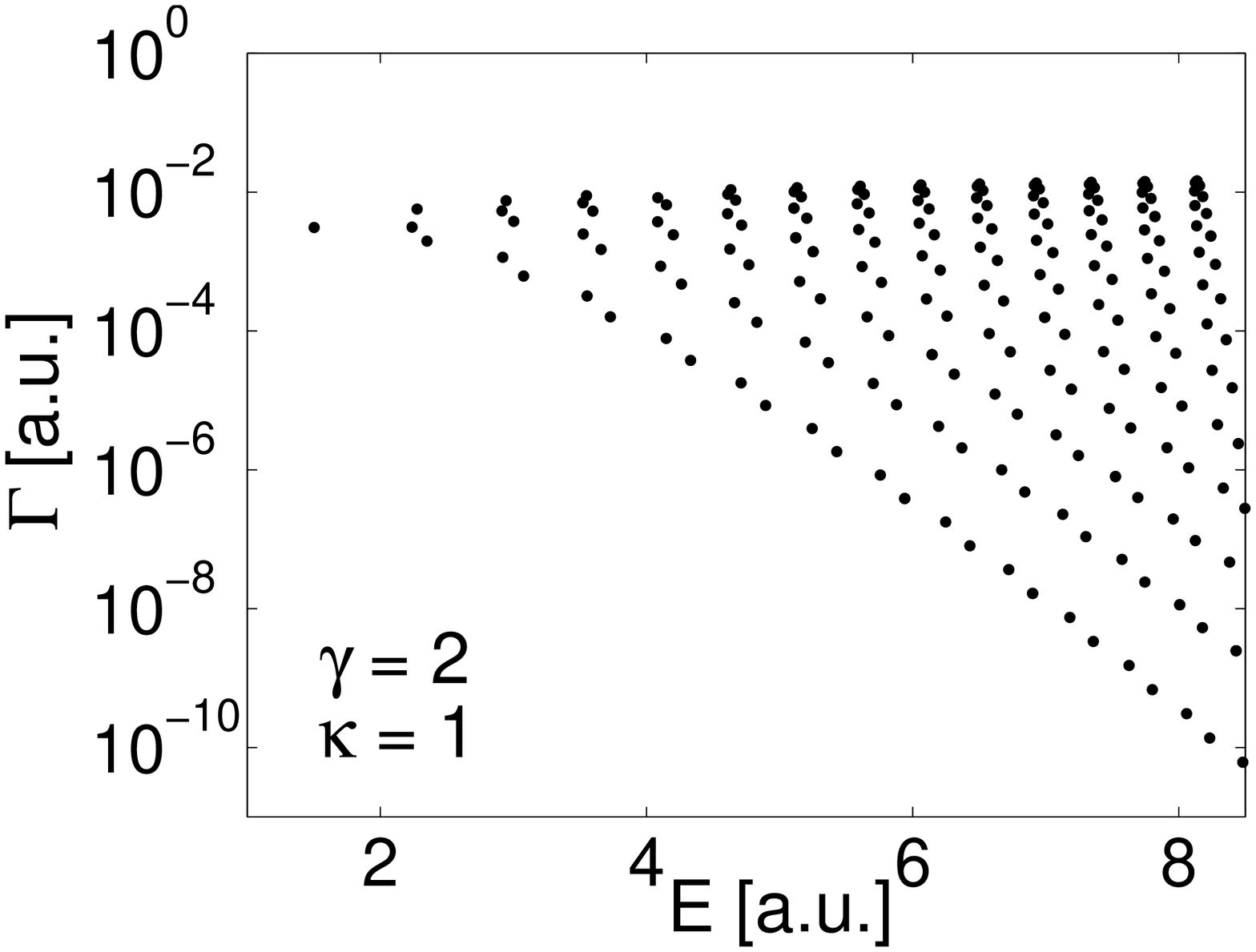}\includegraphics[angle=0,width=7cm]{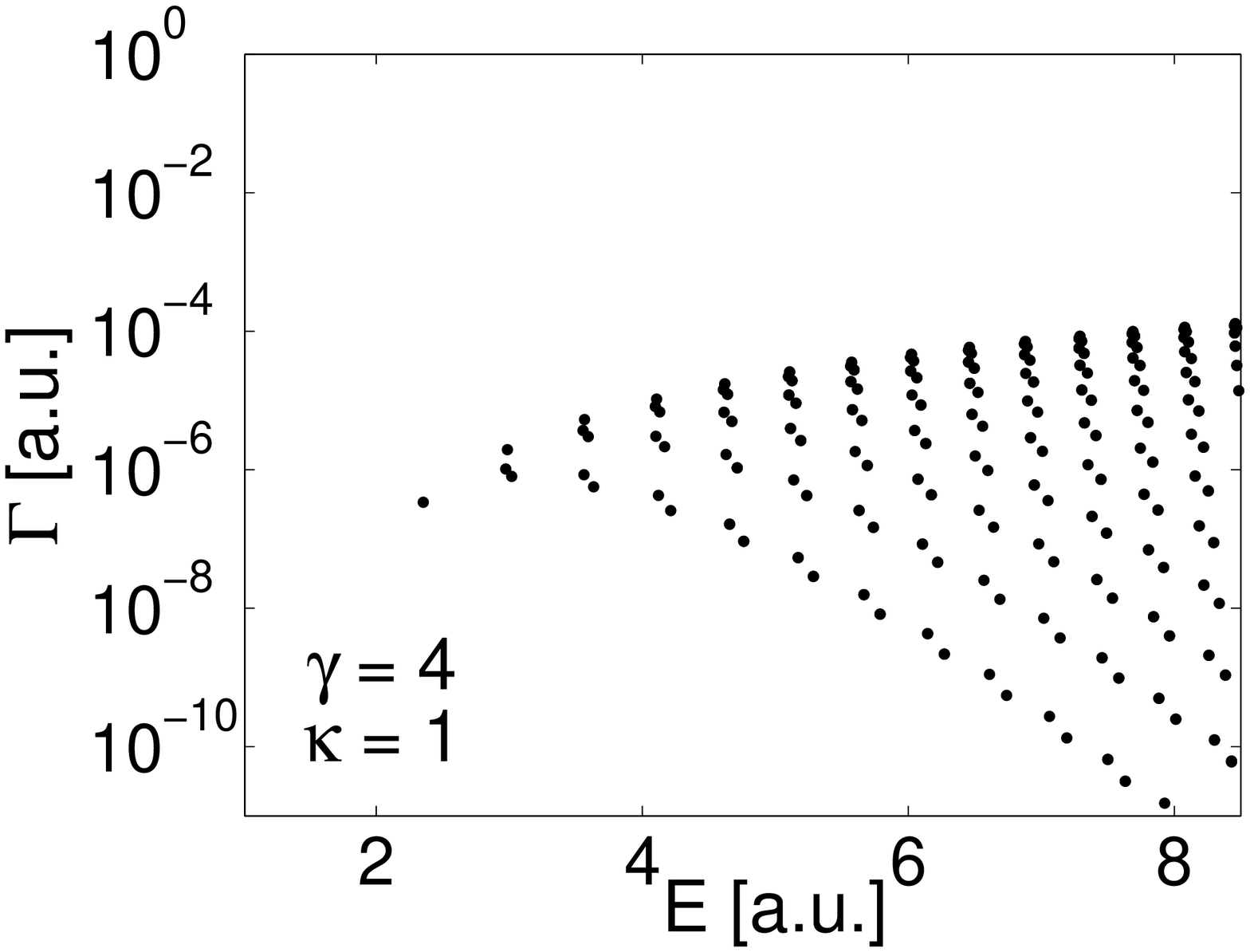}
\caption{Energies $E$ and decay widths $\Gamma$ (logarithmic
scale) for resonances in a magnetic guide for $\kappa=1$. An
increasing Ioffe field strength leads to a global decrease of the
decay widths and therefore has a stabilizing effect. At the same
time the resonance pattern becomes distorted and a regrouping of
the resonances into pairs is observed.} \label{fig:resonances}
\end{figure}

%
In figure \ref{fig:resonances} we present the energies and decay
widths of  the resonances of a spin-$\frac{1}{2}$-particle in the
magnetic guide.
For $\gamma=0$ one observes a regular pattern for the arrangement
of the complex energies, i.e. they are located along lines in the
semi-logarithmic representation. This implies that the decay width
essentially decreases exponentially with increasing energy along
the mentioned lines. The decay width of the lowest (and here also
broadest) resonance is $\Gamma^{\gamma=0}_\text{min}=0.3738$.
Increasing $\gamma$ results in a distortion of the regular
pattern. One observes the formation of vertical lines together
with a regrouping of the resonances into pairs for larger $\gamma$
(see in particular figure \ref{fig:resonances} for $\gamma=4$).
Here states with large decay widths are affected more strongly
than long lived states. Overall we encounter a decrease of the
decay width if $\gamma$ increases. Thus as it is well-known an
additional Ioffe field has a stabilizing effect. For the decay
widths of the lowest resonance state our calculation yields the
series $\Gamma^{\gamma=1}_\text{min}=0.0648$,
$\Gamma^{\gamma=2}_\text{min}=3.08\cdot 10^{-3}$,
$\Gamma^{\gamma=4}_\text{min}=3.38\cdot 10^{-7}$ and
$\Gamma^{\gamma=5}_\text{min}=1.04\cdot 10^{-9}$. These values
suggest an exponential decrease of the decay widths if $\gamma$
increases. This agrees with the results of Sukumar and Brink who
have estimated an exponential increase of lifetimes with growing
Ioffe field strength by an analytical calculation
\cite{Sukumar97}.

\begin{figure}[htb]\center
\includegraphics[angle=0,width=7.3cm]{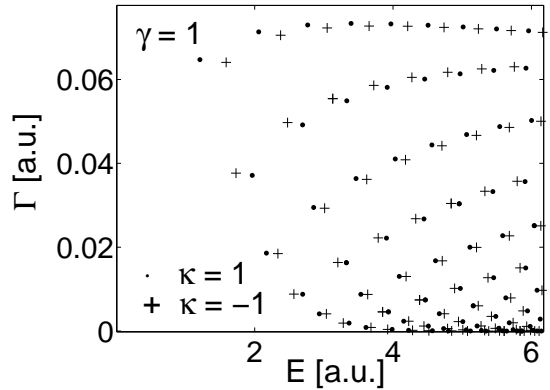}
\caption{Resonance energies and decay widths in the $\kappa=1$- or
$\kappa=-1$-subspaces at $\gamma=1$ (linear scale plot). Due to
the finite Ioffe field formally degenerate pairs split up. States
with negative quantum number $m$ are shifted towards higher
resonance energies and vice versa. The splitting decreases with
decreasing decay width.} \label{fig:resonances_pi_pm1}
\end{figure}

In Sec. \ref{txt:symmetries} we have pointed out the degeneracy of
the $\kappa=1$- and $\kappa=-1$-subspaces. For $\gamma\neq 0$
these degeneracies are expected to split up since the symmetry
properties of the system changes. Figure
\ref{fig:resonances_pi_pm1} shows the resonances for $\gamma=1$.
States belonging to the $\kappa=1$ or $\kappa=-1$ subspaces are
indicated by a dot or cross, respectively.
Each state is accompanied by its degenerate partner for
$\gamma=0$. The respective energy splitting decreases with
decreasing decay width. In the following we will find that a small
decay width is associated to the $\Lambda_z$ eigenvalue property
of a resonance state: States with a small value for $\Gamma$ are
localized far away from the center of the guide. Here the
quadrupole field dominates the homogeneous Ioffe field and
determines the appearence of the spectrum. Therefore these states
become less affected by the Ioffe field which results in a
decrease of the energy splitting.

Let us compare our results to those given by Hinds and Eberlein in
Ref. \cite{Hinds00E}. First of all we remark that we have studied
a significant larger number of resonances.
\begin{table}
\begin{tabular}{l|cc|cc|}
\cline{2-5}
    &\multicolumn{2}{|c|}{ $m=\frac{1}{2}$}   &\multicolumn{2}{|c|}{ $m=\frac{3}{2}$}\\ \cline{2-5}
  &  $E$ & ${\Gamma}$ & $E$& ${\Gamma}$\\ \hline
  \multicolumn{1}{|c|}{}&\textbf{1.2937} & \textbf{0.3738} & \textbf{1.7591} & \textbf{0.1122}\\
  \multicolumn{1}{|c|}{\raisebox{2.5ex}[-2.5ex]{0}}&{1.32} & {0.34} & {1.765} & {0.11}\\\hline
  \multicolumn{1}{|c|}{}&\textbf{2.1404} & \textbf{0.3788} & \textbf{2.5144} & \textbf{0.1444}\\
  \multicolumn{1}{|c|}{\raisebox{2.5ex}[-2.5ex]{1}}&{2.125} & {0.34} & {2.52} & {0.15}\\\hline
  \multicolumn{1}{|c|}{}&\textbf{2.8438} & \textbf{0.3726} & \textbf{3.1711} & \textbf{0.1594}\\
  \multicolumn{1}{|c|}{\raisebox{2.5ex}[-2.5ex]{2}}&{2.81} & {0.34} & {3.17} & {0.18}\\\hline
\end{tabular}
\caption{Comparison between our results (bold face) and those
obtained by Hinds and Eberlein for the first $3$ energetically
lowest resonances in the $m=\frac{1}{2}$ and $m=\frac{3}{2}$
subspaces.} \label{tbl:comparison_to_hinds}
\end{table} In table \ref{tbl:comparison_to_hinds} we have listed the resonance energies and decay
widths of the first $3$ states inside the $m=\frac{1}{2}$ and
$m=\frac{3}{2}$ subspaces. Both results agree within a few
percent. The difference might originate from the complicated
procedure employed by Hinds and Eberlein to locate the resonance
energies \footnote{Complex scaling calculations of Potvliege and
Zehnl\'{e} \cite{Potvliege01} have shown similar discrepancies
with respect to the results given in Ref. \cite{Hinds00}.
Unfortunately the results in both publications are based on the
wrong assumption that the quantum number $m$ is integer-valued
\cite{Hinds00E}.}.
%
\subsection{Dependence of the resonance energies on the eigenvalue of $\Lambda_z$}\label{txt:decay_width__m_eigenvalue}
%
In this section we analyze how the energies and decay widths of
the resonances are related to the eigenvalues of the angular
momentum operator $\Lambda_z$. Although $\Lambda_z$ is a conserved
quantity we have employed a set of basis functions which do not
respect this fact. Thus we need to calculate the matrix element
$m=\left<E,\kappa\right|\Lambda_z\left|E,\kappa\right>$ to receive
the $\Lambda_z$-eigenvalues of the resonances. In order to do this
one has to respect the non-Hermitian character of the Hamiltonian
(\ref{eq:complex_scaled_Hamiltonian}) which requires a complex
symmetric scalar product. For a detailed discussion see Ref.
\cite{Moiseyev98}.
\begin{figure}[htb]\center
\includegraphics[angle=0,width=7cm]{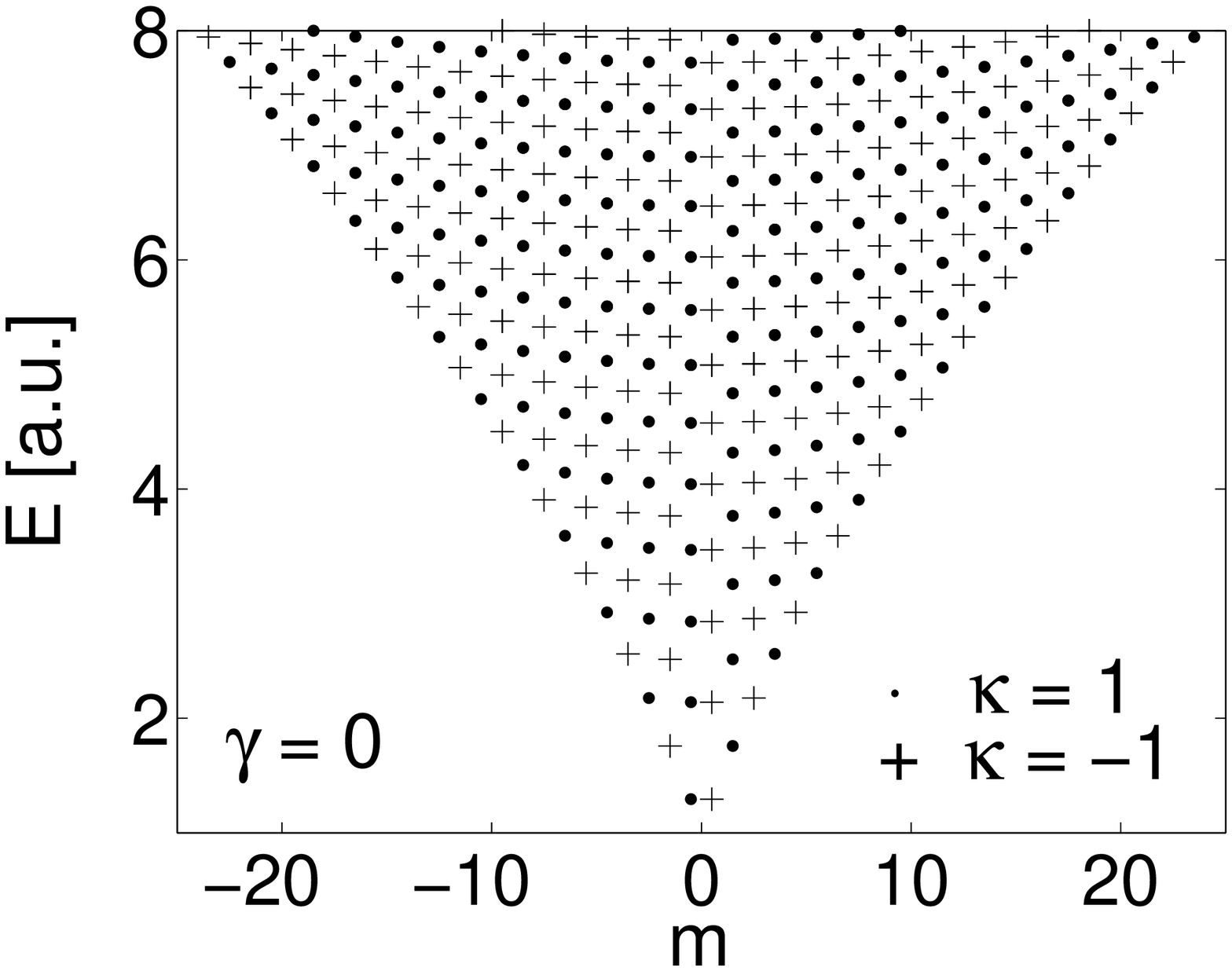}\includegraphics[angle=0,width=7cm]{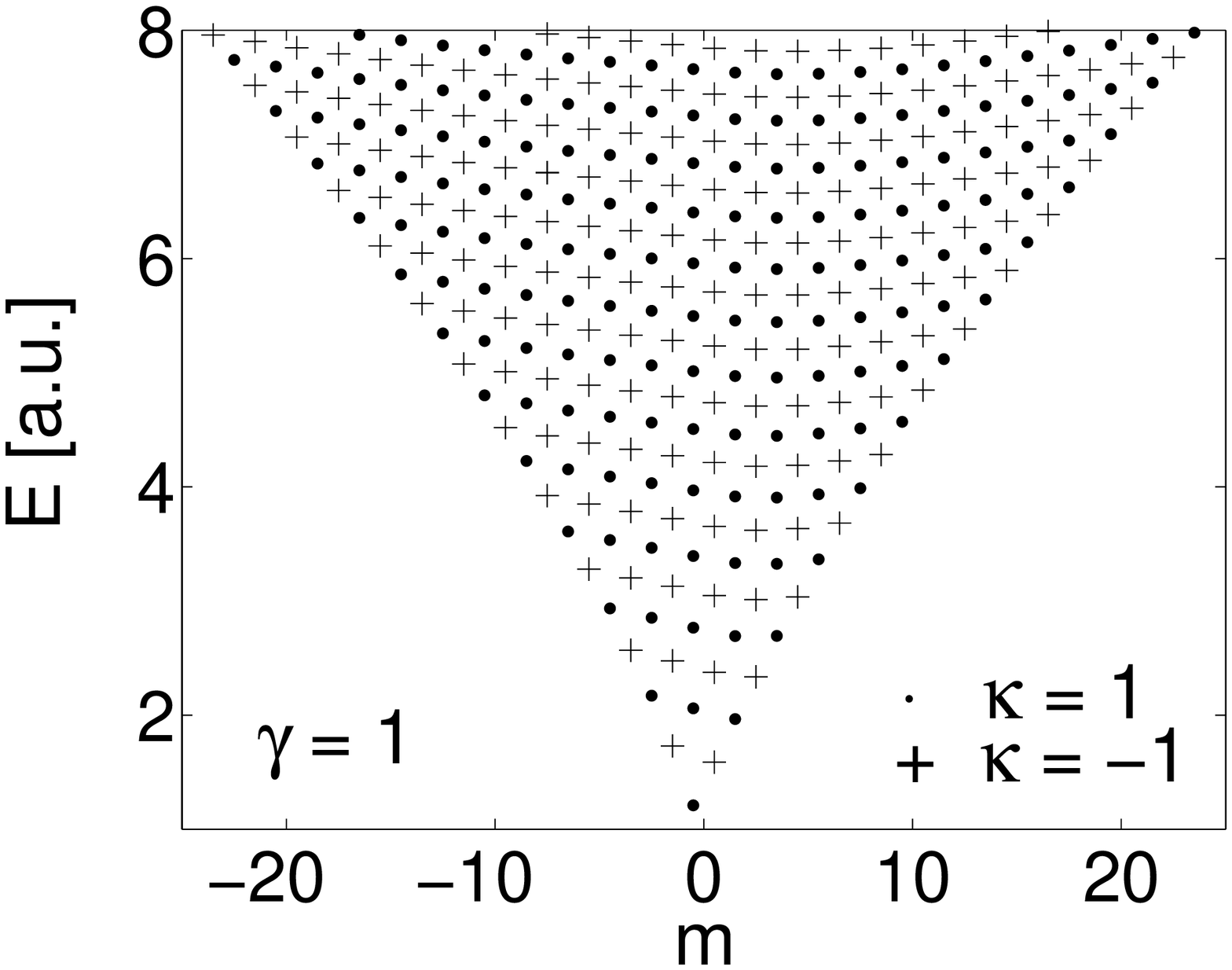}
\caption{Resonance energies and $\Lambda_z$ eigenvalues of
resonances for $\gamma=0$ (left picture) and $\gamma=1$ (right
picture). For $\gamma=0$ the energies of the resonances form a
symmetric pyramid-like pattern. For finite Ioffe field strength
the pattern becomes asymmetric.}
\label{fig:energy_angular_momentum}
\end{figure}
Figure \ref{fig:energy_angular_momentum} depicts the energies $E$
and the $\Lambda_z$ eigenvalue of the resonances. For $\gamma=0$
one observes each state to possess a degenerate counterpart, i.e.
a state with opposite $m$. This results in the formation of a
symmetric pyramid-like distribution where the maximum $\Lambda_z$
eigenvalue depends approximately linear on the resonance energy
$E$. With increasing $\gamma$ the resonance energies in general
are shifted towards larger values (this can hardly be seen in
figure \ref{fig:energy_angular_momentum} for $\gamma=1$). Thereby,
states with positive $m$ acquire a larger energetical shift than
states with negative values of the quantum number $m$. Due to this
asymmetric energy shift the distribution becomes asymmetric (with
respect to $m\rightarrow -m$), too. Here energies of states having
the same $\kappa$ quantum number form continuous, nearly
horizontal, lines whereas we observe the resonance energies for
$\gamma=0$ to be arranged on broken horizontal lines. In
subsection \ref{txt:quasi_bound_gamma} we will show that in the
limit of $\frac{\rho^2}{\gamma^2}\rightarrow 0$ one finds a
pattern of equidistant straight horizontal lines formed by the resonance
energies. Subsequent lines belong to states with opposite
$\Sigma_z$ quantum number. With increasing $m$ values the states
become less affected by the Ioffe field. Their wavefunctions
become located farther away from the center of the guide. In this
region the strength of the linearly increasing quadrupole field
outweighs the effects of the Ioffe field.

\begin{figure}[htb]\center
\includegraphics[angle=0,width=7cm]{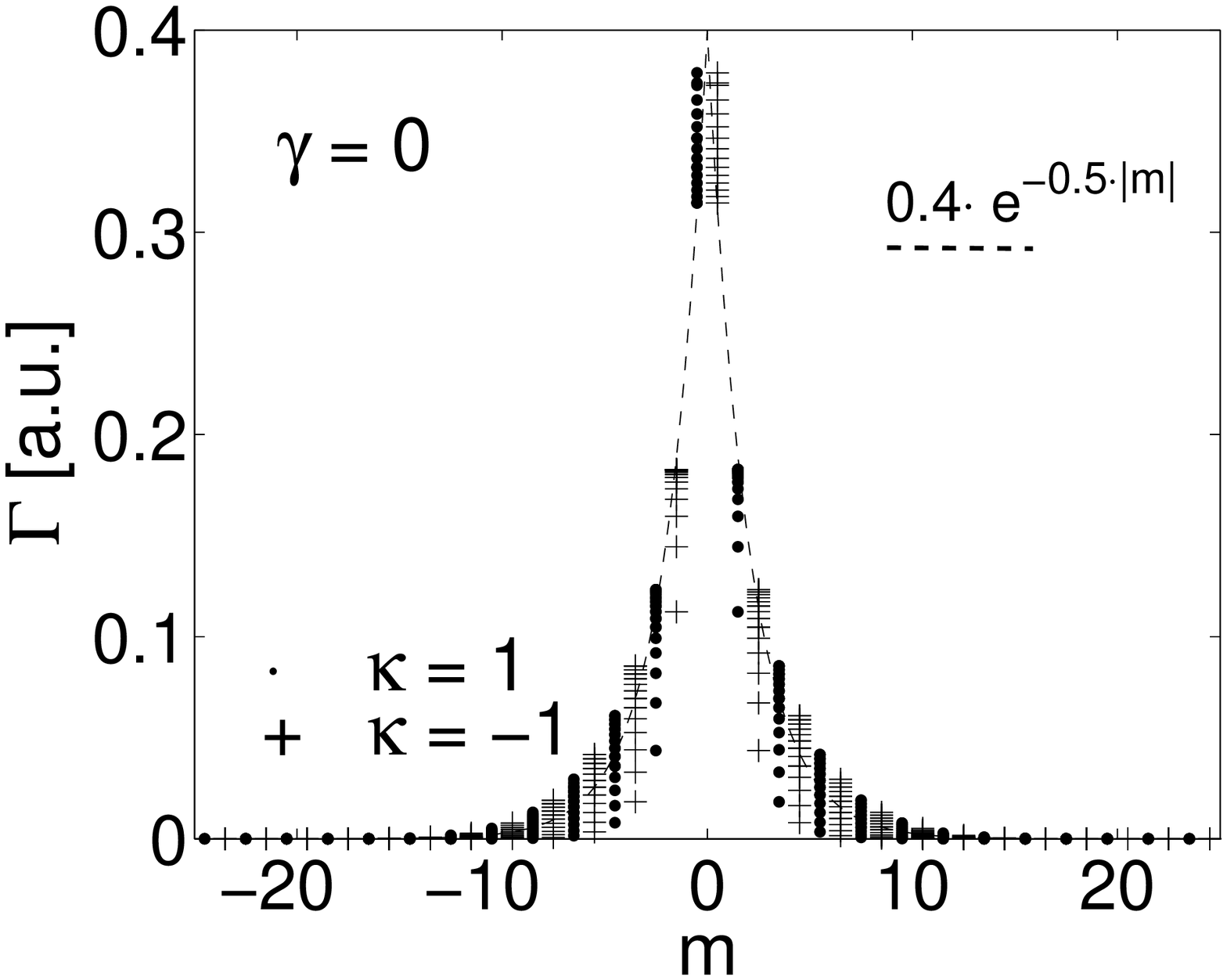}\includegraphics[angle=0,width=7.1cm]{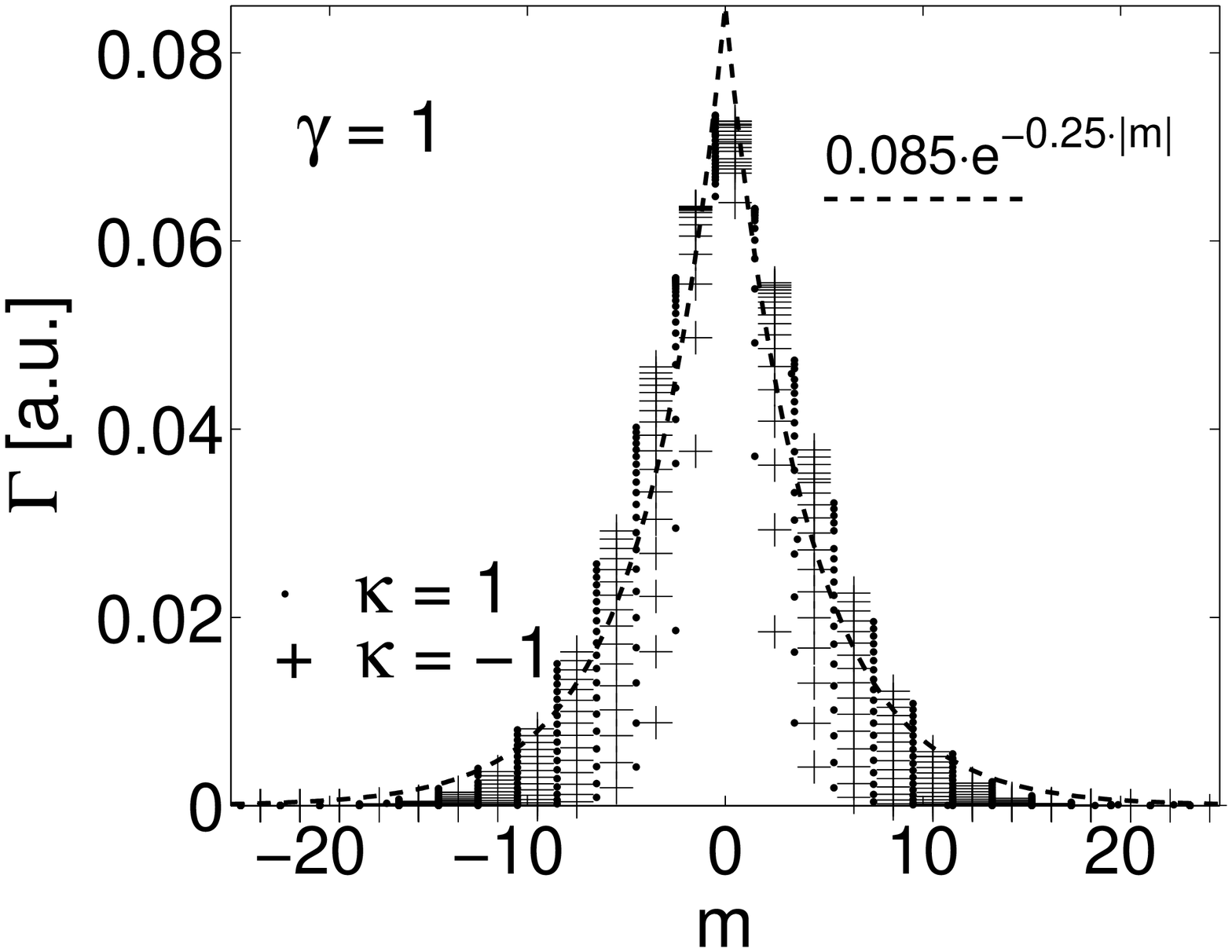}
\caption{Decay widths and $\Lambda_z$ eigenvalues of resonances
for $\gamma=0$ (left picture) and $\gamma=1$ (right picture). The
decay width decreases exponentially with increasing modulus of
$m$. With increasing Ioffe field strength one observes an overall
decrease of the decay widths and a widening of the distribution. }
\label{fig:widths_angular_momentum}
\end{figure}
In figure \ref{fig:widths_angular_momentum} the dependence of the
decay widths on the $\Lambda_z$ eigenvalue is presented. Here we
encounter an exponential decrease of the decay widths with
increasing modulus of $m$. Overall the decay widths decrease if
$\gamma$ increases and at the same time the distribution becomes
broader.
%
\subsection{Quasi-bound states for $\gamma=0$}\label{txt:quasi_bound_states}
%
Although a spin in our field configuration does not possess bound
states there exist resonances with very long lifetimes, i.e. very
small decay widths. The discussion of these quasi-bound states is
the subject of the present section.

Transforming the Hamiltonian (\ref{eq:scaled_Hamiltonian}) to
polar coordinates $(\rho,\phi)$ and exploiting the conservation of
$\Lambda_z$ yields
\begin{eqnarray}
  H_m=\left<m\right|H\left|m\right>=\frac{1}{2}\left[-\frac{\partial^2}{\partial\rho^2}-\frac{1}{\rho}\frac{\partial}{\partial\rho}+\frac{\left(m+\frac{1}{2}\sigma_z\right)^2}{\rho^2}+\rho\sigma_x
  \right].
\end{eqnarray}
Performing the spin-space transformation
\begin{eqnarray}
  U=\frac{1}{\sqrt{2}}\left(%
\begin{array}{cc}
  1 & 1 \\
  1 & -1 \\
\end{array}%
\right)\label{eq:U}
\end{eqnarray}
and introducing the spinor
$\left|\Psi\right>=\rho^{-\frac{1}{2}}\left|\Phi\right>$ the
corresponding Schr\"odinger equation becomes
\begin{eqnarray}
  \frac{1}{2}\left[-\frac{\partial^2}{\partial\rho^2}+\frac{m^2+m\sigma_x}{\rho^2}+\rho\sigma_z
  \right]\left|\Phi\right>=E\left|\Phi\right>.\label{eq:UHU_cylindrical_schroedinger_equation}
\end{eqnarray}
The transformation (\ref{eq:U}) diagonalizes the
$\vec{\mu}{\vec{B}}$ interaction term but leads to off-diagonal
elements in the angular momentum term. In the limit of large $m$
this coupling between the up and down components of
$\left|\Phi\right>$ can be neglected:
\begin{eqnarray}
  \frac{m^2+m\sigma_x}{\rho^2}\rightarrow
  \frac{m^2}{\rho^2}\label{eq:quasi_bound_approximation}
\end{eqnarray}
which  yields
\begin{eqnarray}
 \frac{1}{2}\left[-\frac{\partial^2}{\partial\rho^2}+\frac{m^2}{\rho^2}+\rho\sigma_z
  \right]\left|\Phi\right>=E_{qb}\left|\Phi\right>.\label{eq:quasi_bound_schroedinger_equation}
\end{eqnarray}
By construction this radial Schr\"odinger equation
(\ref{eq:quasi_bound_schroedinger_equation}) does not couple the
up- and down-component of the spinor wavefunction
$\left|\Phi\right>$. The lower component is unbound since the
corresponding effective potential is
$V^-\left(\rho\right)=\frac{m^2}{2\rho^2}-\frac{1}{2}\rho$. In
contrast to this the potential for the upper component is
$V^+\left(\rho\right)=\frac{m^2}{2\rho^2}+\frac{1}{2}\rho$ and
therefore bound solutions are allowed. We will refer to the
corresponding states as the quasi-bound states
$\left|\chi\right>=\left(\chi\left(\rho \right),0\right)^T$ which
are described by the Schr\"odinger equation
\begin{eqnarray}
\frac{1}{2}\left[-\chi^{\prime\prime}+\frac{m^2}{\rho^2}\chi+\rho\chi\right]=E_{qb}\chi.\label{eq:quasi_bound_states}
\end{eqnarray}
Here a prime denotes the derivative with respect to $\rho$. In
order to solve equation (\ref{eq:quasi_bound_states}) we have
utilized the FEMLAB software package wich employs the finite
element method for solving differential equations \footnote{To our
knowledge there is no analytic solution of equation
(\ref{eq:quasi_bound_states}). However, for $\rho\rightarrow 0$
the solutions become cylindrical Bessel-functions whereas for
$\rho\rightarrow\infty $ they behave like Airy-functions.}.

\begin{table}
\begin{tabular}{|c|c||c|c|c|c|c|c|}
\hline

&  & 0 & 1 & 2 & 3 &  4 & 5 \\\hline\hline

&$E$&\textbf{1.2937} &\textbf{2.1404} &\textbf{2.8438}
&\textbf{3.4688} &\textbf{4.0419} & \textbf{4.5769}
\\\hline

 $m=\frac{1}{2}$  &$E_{qb}$&{1.2829} &{2.1287} &{2.8324}
&{3.4579} &{4.0315} &{4.5669}\\\hline

 &\%  &0.83 &0.55 &0.40 &0.31 & 0.26 & 0.22 \\\hline\hline

&$E$&\textbf{3.2660} &\textbf{3.8412} &\textbf{4.3797} &
\textbf{4.8889} & \textbf{5.3741} & \textbf{5.8392}
\\\hline

 $m=\frac{11}{2}$  &$E_{qb}$&{3.2542} &{3.8256} &{4.3612}
&{4.8684} &{5.3522} & {5.8162} \\\hline

  &\%  &0.36 & 0.41 & 0.42 & 0.42 & 0.41 & 0.39 \\\hline\hline

&$E$&\textbf{4.7847} &\textbf{5.2629} &\textbf{5.7230} &
\textbf{6.1673} &\textbf{6.5977} & \textbf{7.0156}\\\hline

 $m=\frac{21}{2}$   &$E_{qb}$&{4.7805} &{5.2577} &{5.7169}
&{6.1602} &{6.5897} & {7.0068}\\\hline

  &\%   & 0.09 & 0.10 & 0.11 &0.12  & 0.12 & 0.13 \\\hline\hline

&$E$&\textbf{6.0956} &\textbf{6.5208} &\textbf{6.9344}
&\textbf{7.3375} &\textbf{7.7311} & \textbf{8.1160}\\\hline

 $m=\frac{31}{2}$   &$E_{qb}$&{6.0934} &{6.5181} &{6.9313}
&{7.3341} &{7.7273} & {8.1118}\\\hline

 &\%    &0.04 & 0.04 & 0.04 & 0.05 & 0.05 & 0.05 \\\hline\hline

&$E$&\textbf{7.2790} &\textbf{7.6689} &\textbf{8.0505} &
\textbf{8.4245} & - & -\\\hline

$m=\frac{41}{2}$   &$E_{qb}$&{7.2774} &{7.6671} &{8.0486} &
{8.4224} & {8.7891}&{9.1492}\\\hline

   &\%  & 0.02 & 0.02 & 0.02 & 0.02 & - & -\\\hline

\end{tabular}
\caption{Comparison of the resonance energies to the approximate
energies $E_{qb}$ obtained from equation
(\ref{eq:quasi_bound_schroedinger_equation}). The first six
resonance energies $E$ for 5 selected values of the quantum number
$m$ are provided. The rows labelled by '\%' show the relative
difference between $E$ and $E_{qb}$ in
percent.}\label{tbl:comparison}
\end{table}
In table \ref{tbl:comparison} the resonance energies $E$ obtained
from the complex scaling calculation are compared to the
approximate energies $E_{qb}$ resulting from equation
(\ref{eq:quasi_bound_schroedinger_equation}). Even for
$m=\frac{1}{2}$ we find a remarkable good agreement between $E$
and $E_{qb}$ although the validity of equation
(\ref{eq:quasi_bound_schroedinger_equation}) is not justified
since the off-diagonal coupling terms are of the same magnitude
than the diagonal ones. With increasing $m$ the discrepancy of $E$
and $E_{qb}$ decreases as suggested by
(\ref{eq:quasi_bound_approximation}). The surprisingly good
quality of the approximation can be explained by looking at how
the bound solution $\left|\chi\right>=\left(\chi\left(\rho
\right),0\right)^T$ and the unbound wavefunction
$\left|\zeta\right>=\left(0,\zeta\left(\rho \right)\right)^T$ are
coupled by the Schr\"odinger equation
(\ref{eq:UHU_cylindrical_schroedinger_equation}):
\begin{eqnarray}
\frac{m}{2}\left<\chi\right|\frac{\sigma_x}{\rho^2}\left|
\zeta\right>=\frac{m}{2}\int d\rho\, \chi^*\left(\rho
\right)\frac{1}{\rho^2}\zeta\left(\rho \right)
\end{eqnarray}
From this expression one recognizes that transitions from the
bound state $\left|\chi\right>$ to the unbound state
$\left|\zeta\right>$ are going to happen essentially at the center
of the guide. However, since $m$ can only adopt half-integer
values the centrifugal barrier in equation
(\ref{eq:quasi_bound_schroedinger_equation}) always persists. Thus
both wavefunctions $\left|\chi\right>$ and $\left|\zeta\right>$
vanish for $\rho\rightarrow 0$. However, this is the only region
where the operator $\frac{1}{\rho^2}$ contributes significantly.
Hence we have $\left<\chi\right|\frac{\sigma_x}{\rho^2}\left|
\zeta\right>\ll 0$ and thus there is only a small coupling between
the bound and unbound solution. Therefore the resonance states are
very well described by the solutions $\left|\chi\right>$ of
equation (\ref{eq:quasi_bound_states}). This also explains the
long lifetime of states with high $m$ quantum numbers. Here the
large angular momentum barrier prevents the coupling between the
bound and unbound channel. The particles are then located between
the classical turning points of the potential
$V_\text{qb}(\rho)=\frac{m^2}{2\rho^2}´+\frac{1}{2}\rho$ at a
distance of approximately $\rho_\text{min}=\sqrt[3]{2\,m^2}$.
\begin{figure}[htb]\center
\includegraphics[angle=0,width=7cm]{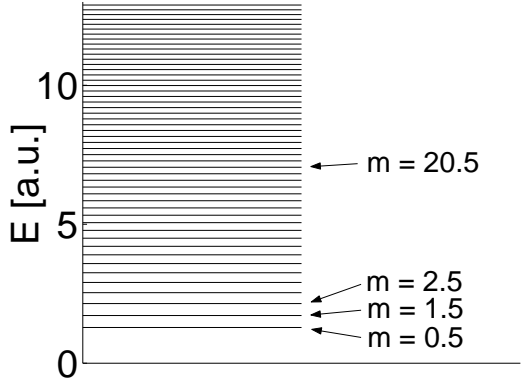}
\caption{Energies $E_\text{lr}$ of the lowest resonances of each
$m$ subspace calculated by using equation (\ref{eq:harm_approx}).
Their accuracy is higher than the resolution of the plot.}
\label{fig:harm_levels}
\end{figure}
Performing a harmonic approximation of the potential $V^+(\rho)$
around its minimum at $\rho_\text{min}$ yields a useful expression
for the energy of the lowest resonance in each $\Lambda_z$
subspace
\begin{eqnarray}
E_\text{lr}=\frac{\sqrt{3}}{2}\left(2\,m^2\right)^{-\frac{1}{3}}+\frac{3}{2}\left(\frac{m}{2}\right)^\frac{2}{3}.
\label{eq:harm_approx}
\end{eqnarray}
This equation provides a very good approximation to the energies.
The calculated resonance energies differ from the energies
$E_\text{lr}$ by at most $0.05 \%$. Figure \ref{fig:harm_levels}
shows the energy spectrum obtained from equation
(\ref{eq:harm_approx}). For large $m$ values the level spacing
scales according to $m^{-\frac{1}{3}}$.

%
\subsection{Quasi-bound states for $\gamma\neq 0$ and adiabatic
approximation}\label{txt:quasi_bound_gamma}
%

In the previous section we have shown the existence of quasi-bound
states in case of a vanishing Ioffe field. We will now show that
similar states exist also if a Ioffe field is present. We start
our investigation with the Hamiltonian
\begin{eqnarray}
  H_m=\left<m\right|H\left|m\right>=\frac{1}{2}\left[-\frac{\partial^2}{\partial\rho^2}-\frac{1}{\rho}\frac{\partial}{\partial\rho}+\frac{\left(m+\frac{1}{2}\sigma_z\right)^2}{\rho^2}+\rho\sigma_x+\gamma\sigma_z
  \right].
\end{eqnarray}
Introducing the spinor wavefunction
$\left|\Psi\right>=\rho^{-\frac{1}{2}}\left|\Phi\right>$ and
applying the unitary transformation
\begin{eqnarray}
  U=\frac{1}{\sqrt{2}}\left(%
\begin{array}{cc}
  \sqrt{1+\alpha} & \sqrt{1-\alpha} \\
  \sqrt{1-\alpha} & -\sqrt{1+\alpha} \\
\end{array}%
\right)\label{eq:U_Ioffe}
\end{eqnarray}
with $\alpha=\frac{\gamma}{\sqrt{\gamma^2+\rho^2}}$ the
corresponding Schr\"odinger equation becomes
\begin{eqnarray}
  \frac{1}{2}\left[-\frac{\partial^2}{\partial\rho^2}+\frac{m^2}{\rho^2}
  +\frac{m\alpha}{\rho^2}\sigma_z+\frac{\gamma}{\alpha}\sigma_z
  +\frac{\alpha^4}{4\gamma^2}-i\left(\frac{\alpha^2}{\gamma}\frac{\partial}{\partial\rho}-\rho\frac{\alpha^4}{\gamma^3}\right)\sigma_y+\frac{m\alpha}{\gamma\rho}\sigma_x\right]\left|\Phi\right>
  =E\left|\Phi\right>\label{eq:transformeg_seq_ioffe}
\end{eqnarray}
Like without Ioffe field the unitarian (\ref{eq:U_Ioffe})
transforms the $\vec{\mu}\vec{B}$-coupling term into a diagonal
form. However, unlike the transformation (\ref{eq:U}) $U$ in
equation (\ref{eq:U_Ioffe}) depends explicitly on the coordinate
$\rho$. Therefore the transformation of the derivative results in
additional terms:
\begin{eqnarray}
U^+\frac{\partial^2}{\partial\rho^2}U&=&
U^+U''+2U^+U'\frac{\partial}{\partial\rho}+\frac{\partial^2}{\partial\rho^2}.\label{eq:derivative_transform}
\end{eqnarray}

Neglecting the off-diagonal coupling terms in equation
(\ref{eq:transformeg_seq_ioffe}) yields:
\begin{eqnarray}
  \frac{1}{2}\left[-\frac{\partial^2}{\partial\rho^2}+\frac{m^2}{\rho^2}
  +\frac{m\alpha}{\rho^2}\sigma_z+\frac{\gamma}{\alpha}\sigma_z
  +\frac{\alpha^4}{4\gamma^2}\right]\left|\Phi\right>
  =E_{qb}\left|\Phi\right>
\end{eqnarray}
Here similar to equation
(\ref{eq:quasi_bound_schroedinger_equation}) only the upper
component of the spinor $\left|\Phi\right>$ is bound. It obeys the
Schr\"odinger equation
\begin{eqnarray}
  \frac{1}{2}\left[-\frac{\partial^2}{\partial\rho^2}+\frac{m^2}{\rho^2}+\sqrt{\gamma^2+\rho^2}
  +\frac{m\gamma}{\rho^2\sqrt{\gamma^2+\rho^2}}
  +\frac{\gamma^2}{4\left(\gamma^2+\rho^2\right)^2}\right]\chi
  =E_{qb}\chi.\label{eq:quasi_bound_schroedinger_equation_gamma}
\end{eqnarray}

\begin{table}
\begin{tabular}{|c|c||c|c|c|c|c|c|}
\hline

&  & 0 & 1 & 2 & 3 &  4 & 5 \\\hline\hline

&$E$&\textbf{2.8163} &\textbf{3.3978} &\textbf{3.9412}
&\textbf{4.4547} &\textbf{4.9437} &\textbf{5.4123}
\\\hline

 $m=-\frac{1}{2}$  &$E_{qb}$&2.8471 & 3.4272 & 3.9691 & 4.4813 & 4.9692 & 5.4369 \\\hline

 &\%  &1.08  &  0.86  &  0.70  &  0.59  &  0.51  &  0.45
 \\\hline\hline

 $l=0$  &$E_{ad}$&2.8387 & 3.4205 & 3.9635 & 4.4764 & 4.9649 & 5.4329
 \\\hline

 &\%  &0.79  &  0.66  &  0.56  &  0.48  &  0.43  &  0.38
 \\\hline
\hline

&$E$&\textbf{4.4663} & \textbf{4.953} &  \textbf{5.4199} &
\textbf{5.8696} & \textbf{6.3045} & \textbf{6.7262}
\\\hline

 $m=\frac{11}{2}$  &$E_{qb}$&4.4648 & 4.9514 & 5.4180  & 5.8677 & 6.3024 & 6.7240  \\\hline

  &\%  &0.03 &  0.03 &  0.04 &  0.03 &  0.03 &  0.03
 \\\hline\hline

 $l=6$  &$E_{ad}$&4.4957 & 4.9797 & 5.4442 & 5.8921 & 6.3253 & 6.7457
  \\\hline

  &\%  &0.65  &  0.54  &  0.45  &  0.38  &  0.33  &  0.29

 \\\hline\hline

&$E$&\textbf{6.8534} & \textbf{7.2540} &  \textbf{7.6453} &
\textbf{8.0280} & \textbf{8.4028} & \textbf{8.7703}
\\\hline

 $m=\frac{31}{2}$   &$E_{qb}$&6.8523 & 7.2528 & 7.6440 &  8.0267 & 8.4014 & 8.7688
\\\hline

 &\%    &0.02 &  0.02 &  0.02 &  0.02 &  0.02 &  0.02
 \\\hline\hline

  $l=16$   &$E_{ad}$&6.9044 & 7.3027 & 7.6918 & 8.0725 & 8.4455 & 8.8114

\\\hline

 &\%    &0.74  &  0.67 &   0.60 &   0.55  &  0.51  &  0.47
 \\\hline

\end{tabular}
\caption{Comparison of the resonance energies to the approximate
energies $E_{qb}$ and adiabatic energies $E_{ad}$ for $\gamma=5$.
The first six resonance energies $E$ for 5 selected values of the
$m$ quantum number are given. The rows labelled by '\%' show the
relative difference between $E$ and $E_{qb}$ or $E_{ad}$,
respectively, in percent.}\label{tbl:comparison_gamma}
\end{table}
Table \ref{tbl:comparison_gamma} compares the resonance energies
$E$ obtained for $\gamma=5$ to the eigenvalues $E_{qb}$ calculated
by solving the scalar radial Schr\"odinger equation
(\ref{eq:quasi_bound_schroedinger_equation_gamma}). Apart from
$m=-0.5$ the agreement is excellent with discrepancies lower than
0.05 \%. This is again the result of a localization of the
particle wavefunction away from center of the guide which is the
only region where the off-diagonal coupling terms are remarkable.
However, for $m=-0.5$ and $\gamma=5$ the effective potential of
the Schr\"odinger equation
(\ref{eq:quasi_bound_schroedinger_equation_gamma}) does not
possess a centrifugal barrier. Here the wavefunction has non-zero
contributions in the vicinity of the center of the guide. Hence
the off-diagonal coupling terms of equation
(\ref{eq:transformeg_seq_ioffe}) become important. Considering the
fact that the equation
(\ref{eq:quasi_bound_schroedinger_equation_gamma}) here becomes
certainly invalid the energies $E$ and $E_{qb}$ still agree
surprisingly well (see table \ref{tbl:comparison_gamma}).

We now compare the results of the approximate Schr\"odinger
equation (\ref{eq:quasi_bound_schroedinger_equation_gamma}) to
those one would obtain within the so called adiabatic
approximation. In this picture one assumes the projection of the
atomic spin onto the local direction of the magnetic field to be
conserved. Thus the coupling of the magnetic moment to the field
reduces to $g\mu_B m_S\left|\vec{B}\right|$ with $m_S$ being the
projection of the spin onto the local field direction. In case of
a spin-$\frac{1}{2}$-particle in the magnetic guide the
corresponding Hamiltonian becomes
\begin{eqnarray}
H_{ad}=\frac{1}{2}\left[p_x^2+p_y^2\pm\sqrt{\gamma^2+x^2+y^2}\right]\label{eq:H_ad}
\end{eqnarray}
having employed scaled coordinates (see Sec. \ref{txt:system}).
Considering only the positive sign (which allows bound solutions)
and introducing the wavefunction
$\left|\Psi_{ad}\right>=\rho^{-\frac{1}{2}}\left|\Phi_{ad}\right>$
the corresponding Schr\"odinger equation becomes
\begin{eqnarray}
  \frac{1}{2}\left[-\frac{\partial^2}{\partial
  \rho^2}+\frac{l^2-\frac{1}{4}}{\rho^2}+\sqrt{\gamma^2+\rho^2}\right]\left|\Phi_{ad}\right>=E_{ad}\left|\Phi_{ad}\right>.\label{eq:SEQ_ad}
\end{eqnarray}
Here $l$ is the quantum number of the operator $L_z$ which is
conserved due to the rotational invariance of the system around
the $z$-axis. Note that unlike $m$ the quantum number $l$ is
integer-valued. Table \ref{tbl:comparison_gamma} shows a
comparison of the adiabatic eigenvalues $E_{ad}$ to the exact
resonance energies as well as the energies of the quasi-bound
states $E_{qb}$. The quantum numbers $l$ and $m$ are chosen such
that $l=m+\frac{1}{2}$. Only for $l=0$ or $m=-\frac{1}{2}$ the
adiabatic energies $E_{ad}$ are in better agreement to the exact
ones than the quasi-bound energies $E_{qb}$. One has to note that
in contrast to $E_{ad}$ the energies $E_{qb}$ become exact in the
limit of high $m$ quantum numbers. Moreover we have to emphasize
that only the Schr\"odinger equation
(\ref{eq:quasi_bound_schroedinger_equation_gamma}) yielding the
quasi-bound states reproduces the correct degeneracies of the
system. In contrast to this the corresponding adiabatic equation
shows a two-fold degeneracy of the states $\left|l\right>$ and
$\left|-l\right>$ for any value of $\gamma$.
\begin{figure}[htb]\center
\includegraphics[angle=0,width=7cm]{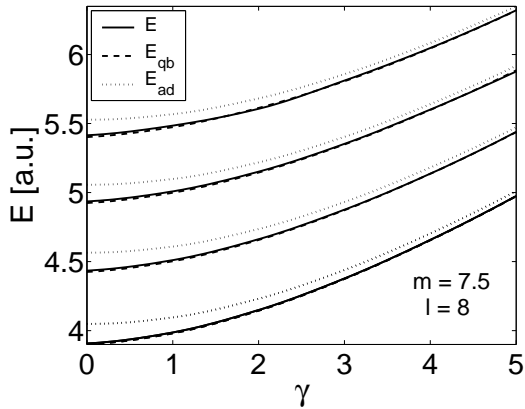}
\caption{Comparison of the quasi-bound energies $E_{qb}$ and
adiabatic energies $E_{ad}$ to the exact energies $E$. The figure
shows the energies of the lowest four states in the
$m=\frac{15}{2}$ and $l=8$ subspace, respectively. The discrepancy
between $E$ and $E_{qb}$ is hardly visible throughout the complete
$\gamma$-interval shown.} \label{fig:comparison}
\end{figure}
We now investigate how the different approximations perform for
different $\gamma$ values. In figure (\ref{fig:comparison}) the
energies obtained from each of the three methods (exact,
quasi-bound, adiabatic) are depicted for the 4 energetically
lowest resonances in the $m=\frac{15}{2}$ and $l=8$ subspace,
respectively. One observes a remarkable agreement of $E$ and
$E_{qb}$ throughout the complete $\gamma$-interval. In contrast to
that severe discrepancies between the adiabatic and exact energies
are revealed for small values of $\gamma$. This shows the
extremely good performance of the quasi-bound approximation
independently of the value of $\gamma$. This makes equation
(\ref{eq:quasi_bound_schroedinger_equation_gamma}) the correct
choice to calculate approximate eigenvalues.

Figure (\ref{fig:comparison}) also shows that if $\gamma$ becomes
large the adiabatic approximation begins to perform well. In the
limit of $\frac{\rho^2}{\gamma^2}\ll 1$ equation (\ref{eq:SEQ_ad})
can be further simplified. A series expansion of the potential
term up to first order in $\frac{\rho^2}{\gamma^2}$ yields
\begin{eqnarray}
  \frac{1}{2}\left[-\frac{\partial^2}{\partial
  \rho^2}+\frac{l^2-\frac{1}{4}}{\rho^2}+\frac{\rho^2}{2\gamma}+\gamma\right]\left|\Phi_{ad}\right>=E_{ad}\left|\Phi_{ad}\right>.\label{eq:SEQ_ad_taylor}
\end{eqnarray}
This is the Schr\"odinger equation of a radial harmonic
oscillator. The corresponding eigenenergies are given through
\begin{eqnarray}
E_{n
l}=\frac{\gamma}{2}+\sqrt{\frac{1}{2\gamma}}\left(2n+\left|l\right|+1\right).
\end{eqnarray}
Introducing the frequency $\omega=\sqrt{\frac{1}{2\gamma}}$
together with the substitution $l=m+\frac{1}{2}$ one arrives at
the formula
\begin{eqnarray}
E_{n
m}=\frac{\gamma}{2}+\omega\left(2n+\left|m+\frac{1}{2}\right|+1\right).\label{eq:ad_approx_harmonic}
\end{eqnarray}
\begin{figure}[htb]\center
\includegraphics[angle=0,width=7cm]{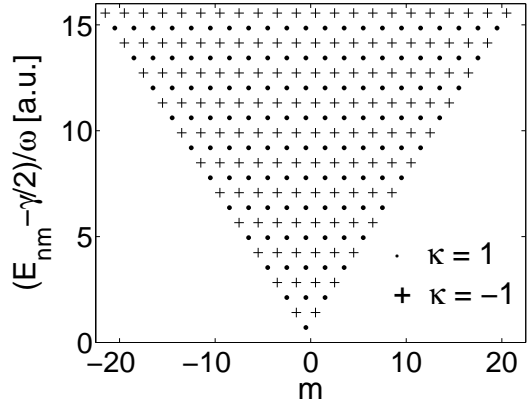}
\caption{Graphical representation of the eigenenergies
(\ref{eq:ad_approx_harmonic}). The values of the quantum number
$\kappa$ are calculated using equation (\ref{eq:kappa_m_mapping}).
A pattern of equidistant horizontal lines with alternating
$\kappa$ values is formed.}
\label{fig:ad_approx_harmonic_energies}
\end{figure}
Figure \ref{fig:ad_approx_harmonic_energies} shows a plot of the
energies $E_{nm}$. The corresponding
$\Sigma_z$-eigenvalues are calculated using equation
(\ref{eq:kappa_m_mapping}). The resultant pattern is similar to
the one we have already observed in figure
\ref{fig:energy_angular_momentum}b. We find alternating
equidistant horizontal lines of energies belonging to states of
the two different $\Sigma_z$-subspaces.
%
\subsection{Resonance energies and decay widths of $^6\text{Li}$ in the $2\,^2\text{S}_\frac{1}{2}\,,\,\text{F}=\frac{1}{2}$ hyperfine ground state
}\label{txt:lithium}
%

In order to experimentally prepare a magnetic guide one can
superimpose the magnetic field of a current carrying wire by a
so-called bias field oriented perpendicular to the current flow
\cite{Folman02}. The resulting magnetic field possesses a line of
zero field strength at a distance $\rho_0=\frac{\mu_0 I}{2 \pi
B_B}$ parallel to the wire. A series expansion of the field around
$\rho_0$ yields
\begin{eqnarray}
\vec{B}\approx\frac{B_B}{\rho_0}\left(%
\begin{array}{c}
  x \\
  -y \\
  0 \\
\end{array}%
\right)+\frac{B_B}{\sqrt{2}\rho_0^2}\left(%
\begin{array}{c}
  -x^2+2xy+y^2 \\
  x^2+2xy-y^2 \\
  0 \\
\end{array}%
\right)+\frac{B_B}{\rho_0^3}\left(%
\begin{array}{c}
  y\left(y^2-3x^2\right) \\
  -x\left(x^2-3y^2\right) \\
  0 \\
\end{array}%
\right).
\end{eqnarray}
These are the quadrupolar, hexapolar and octopolar components of
the field. As long as $\rho_0 \gg 1$ one can neglect the higher
order terms ending up with equation (\ref{eq:sideguide_field}) for
$B=0$. The gradient $b=\frac{2 \pi B^2_B}{\mu_0 I} $ is fully
determined by the strength $B_B$ of the bias field and the current
$I$.

For our discussion we choose the experimental parameters $I=2\,A$
and $B_B=10^{-2} T$. These are typically achievable values which
are used for realizing magnetic microtraps. For this setup one
finds the line of vanishing field strength at a distance
$\rho_0=40 \mu m$ above the wire. The gradient $b$ evaluates to
$b=250 \frac{T}{m}$. For the Land\'{e}-factor of $^6\text{Li}$ in
its ground state
($2\,^2\text{S}_\frac{1}{2},\text{F}=\frac{1}{2}$) one obtains
$g=|g_F|=|-\frac{2}{3}|$. Together with the mass $M=10964.67 m_e$
the energy and the length scale become $\frac{1}{M}\left(\frac{b
gM}{2}\right)^{\frac{2}{3}}=0.864\cdot 10^{-9} eV$ and
$\left(\frac{b gM}{2}\right)^{-\frac{1}{3}}=89.63 nm$,
respectively. The resonance energy of the ground state is $E=1.12
neV$ which corresponds to a temperature of $12.97 \mu K$. This
temperature regime is certainly accessible by todays cold atom
experiments. Correspondingly the transition frequencies to excited
states lie in the $\text{MHz}$ regime. The lifetime of the ground
state evaluates to $2.04 \mu s$. Under consideration of the
exponential scaling with increasing angular momentum almost
arbitrarily long lifetimes can be achieved by preparing the atoms
in high $\Lambda_z$ eigenstate, e.g. the minimum lifetime for a
state with $|m|=\frac{41}{2}$ is $46.61 s$.

%
\section{Conclusion and Outlook}
%
We have investigated the motion of a neutral
spin-$\frac{1}{2}$-fermions in a magnetic quadrupole guide. The
impact of an additionally applied homogeneous Ioffe field is also
studied. Introducing a canonical scaling transformation of the
phase space coordinates we have derived an effective
two-dimensional Hamiltonian depending on a single parameter
$\gamma$. The energies and decay widths of resonance states of the
Schr\"odinger equation have been calculated by employing the
complex scaling method. Utilizing a 2-dimensional harmonic
oscillator basis we were able to converge hundreds of resonance
states.

The analysis of the underlying Hamiltonian revealed a large number
of symmetries. In the absence of a homogeneous Ioffe field we have
found $15$ discrete symmetries of both unitary and anti-unitary
character in addition to the conserved quantity
$\Lambda_z=L_z-S_z$. A deeper investigation of the underlying
symmetry group revealed a two-fold degeneracy of any energy level.
If a Ioffe field is applied $\Lambda_z$ is still conserved but
only 7 discrete symmetries remain. Due to the altered symmetry
group the degeneracies are lifted.

We have calculated the resonance energies for several values of
the parameter $\gamma$. A comparison to the values obtained by
Hinds and Eberlein (Ref. \cite{Hinds00E}) has been performed. For
$\gamma=0$ (vanishing Ioffe field) the resonance energies and
decay widths form a regular pattern in the $E-\Gamma$ plane.
Increasing the Ioffe field leads to a distorted distribution and
the decay widths are pushed towards larger values. Thus the
stability of the resonance states increases with the Ioffe field
strength. An analysis of the lowest resonance has shown an
exponential increase of lifetimes which agrees with results
obtained by others \cite{Sukumar97}. Furthermore we could show an
exponentially increasing lifetime with increasing modulus of the
$m$ quantum number. Apparently, with increasing $|m|$ the
wavefunctions become localized farther from the center of the
guide where transitions to continuum states are induced.

We could show the existence of so called quasi-bound states which
can be described by a scalar radial Schr\"odinger equation. The
approximate eigenenergies agree very well with the resonance
energies obtained from the complex scaling calculation and become
exact in the limit of high $m$ quantum numbers. But even for low
angular momenta an astonishing agreement could be observed. For
$\gamma=0$ (without Ioffe field) this is due to the fact that $m$
is half-integer valued which leads to a non-vanishing angular
momentum barrier. This prevents the particle from entering the
center of the guide. However, the coupling matrix element to the
unbound states does only assume significant values for
$\rho\rightarrow 0$ i.e. the corresponding transitions are
strongly inhibited. For $\gamma=0$ we have also calculated an
analytic expression for the ground state energy in each $m$
subspace. For $\gamma\neq 0$ the quasi-bound energies are compared
to those obtained from the so-called adiabatic approximations. We
have shown that our approach is in general more accurate and
reproduces in particular underlying degeneracies.

The results have been applied to the case of $^6\text{Li}$ in the
$F=\frac{1}{2}$ state. Here we have considered a magnetic guide
generated by a current carrying wire together with a homogeneous
bias field. We have shown that for typical experimental parameter
values the ground state energy corresponds to a temperature of a
few micro-Kelvin. The lifetime of the resonance states can be
extended up to minutes if the atoms are prepared in a sufficient
high angular momentum state.

\end{document}